\documentclass[
]{ceurart}

\usepackage{listings}
\begin{document}

\copyrightyear{2022}
\copyrightclause{Copyright for this paper by its authors.
  Use permitted under Creative Commons License Attribution 4.0
  International (CC BY 4.0).}

\conference{HHAI'25: The 4th International Conference Series on Hybrid Human-Artificial Intelligence, workshop Mind the AI-GAP 2025: Co-Designing Socio-Technical Systems (June 9-13, 2025 in Pisa, Italy)}

\title{When no one shows up (at first) : Navigating the uncertainties of participatory workshops in interdisciplinary research}






\author{Monique Munarini}
\address{University of Pisa, Largo Bruno Pontecorvo, 3, Pisa, Italy}

\begin{abstract} 
This reflective paper explores often-unspoken challenges of designing and facilitating co-design and participatory workshops, offering practical strategies for early career researchers (ECRs) navigating these methods. Drawing from personal experience conducting a series of workshops titled “How to Think About Equity in the AI Ecosystem?”, it follows the full arc of the workshop experience—from conceptualization and activity planning to participant recruitment and facilitation—offering a grounded account of what happens when participation doesn’t go as expected. The paper examines the methodological challenges of engaging non-expert participants, particularly when operating without institutional support, financial incentives, or integration into larger events. Despite initial difficulties—such as low attendance—the workshop fostered rich discussions among a demographically diverse group and ultimately led to one participant volunteering to co-facilitate a subsequent session. This transition from participant to co-facilitator exemplifies the redistribution of epistemic authority, positioning lived experience as central to research and engagement practices. By reframing perceived failure as a productive site of learning, the paper offers practical strategies for ECRs working across disciplines who often navigate unfamiliar methodological terrains contributing to broader conversations on the realities of doing interdisciplinary, participatory work in practice.

\end{abstract}


\maketitle

\section{Introduction}

Participatory research offers an opportunity to shift the power dynamics of knowledge production by bringing affected communities into the design and evaluation of reliable AI systems \cite{Birhanepower2people}. Yet, the practice of organizing such spaces—particularly as an early career researcher navigating interdisciplinary terrain—rarely goes as smoothly as the literature implies. This paper reflects on the design, delivery, and lessons learned from a series of participatory workshops titled “How to Think About Equity in the AI Ecosystem?”. Developed as part of a broader inquiry into equitable AI governance \cite{munarini2025practicing}, the workshops were designed to bring together two often disconnected groups: AI practitioners and those affected by AI systems, with a particular focus on youth—who are positioned to experience the long-term societal impacts of algorithmic decision-making. 

The workshops focused on AI systems in recruitment because access to employment is foundational to economic security, social inclusion, and the protection of human rights. Marginalised groups—such as racialised individuals, migrants, people with disabilities, and gender-diverse persons—often face systemic barriers to labour market access. The increasing adoption of AI-based tools in hiring, while often framed as efficiency-driven or bias-reducing, has revealed significant risks of deepening these inequalities. Notable cases include Amazon’s use of a hiring algorithm that systematically downgraded applications from women due to historical data bias \cite{dastin2022amazon}, and job advertisement algorithms that failed to show employment opportunities to certain groups—such as women or older candidates—because of biased optimisation for engagement metrics \cite{ali2019discrimination}.
Across three sessions, the workshops aimed to explore how equity could be meaningfully embedded into the development of AI systems, particularly in high-stakes domains such as automated hiring. Each session included interactive activities, co-design exercises, and a case study on recruitment tools to support participants in understanding and critiquing how equity operate within these systems. While the workshop for AI practitioners saw high engagement with 14 participants, the sessions involving youth—representing affected groups—revealed the often unpredictable nature of public engagement. One session began with only a single participant in the room and was nearly cancelled before three additional attendees arrived. Another session had a stronger turnout of 11 participants but presented its own logistical and facilitation challenges.

These moments of uncertainty are rarely acknowledged in academic reporting, yet they carry important methodological insights—especially for early career researchers with limited funding, institutional support, or disciplinary training in participatory methods. This paper offers a reflection on the process of developing and hosting participatory workshops, with a particular focus on the session with low attendance. Rather than framing the experience as a failure, it is reframed as a valuable point of learning: an invitation to rethink how we define success in participatory work and how we prepare for unpredictability.

This paper offers a reflection-in-practice, structured around the key stages of workshop development and facilitation. It focuses on the \textit{process} of implementing participatory workshops as a methodological approach—rather than reporting full data analysis or outcomes. While insights from participant discussions are referenced to illustrate key points, the systematic analysis of workshop data and indicators co-designed during the sessions will be presented in a separate, forthcoming publication. First, this paper introduces the motivation and structure of the workshop series, detailing the design process and the specific goals of engaging both AI practitioners and affected groups—particularly youth. The following section offers an observational narrative of the most unpredictable session—where attendance was nearly zero at first—highlighting practical adaptations made in real time. Building on this, the paper reflects on what worked, what did not, and how we could reframe such experiences as methodologically generative rather than failed. The final section offers practical recommendations for early-career researchers undertaking similar work under resource constraints, arguing that small-scale participatory practices still hold valuable lessons.

\section{Planning the workshop}

The series of workshops titled “How to think about equity in the AI ecosystem?” were designed to explore how equity could be integrated into AI systems, particularly within automated hiring. The final goal is to operationalise the equity definition \textit{'Providing meaningful access to the necessary resources for individuals who need it to belong to a community'} proposed by \cite{munarini2024equitable}.This definition draws on feminist theory and the design justice approach, emphasizing not equality of input, but rather redistributive measures tailored to historical and structural exclusions. In this context, “belonging” refers not merely to inclusion, but to a sense of recognition, participation, and influence in sociotechnical decision-making spaces. The title refers to AI ecosystems as developed by Stahl \cite{stahl2021artificial} that AI systems are developed and deployed within multiple relational fields of knowledge, in analogy of a biological ecosystem. The workshop draws from participatory design traditions that emphasise mutual learning, horizontal power dynamics, and situated knowledges, treating participants not as users or test subjects but as co-creators of socio-technical imaginaries \cite{fine2004critical,collins2000black}.
This research chose to embrace the challenge of working with the “messy middle” as defined by \cite{Patelframework} as a combination between the perspectives of AI practitioners and affected communities. This paper will focus on one of the workshops in this series.
This specific session targeted youth, broadly defined as individuals aged approximately 18 to 35, encompassing millennials and Generation Z. This age group was chosen because they are poised to experience the long-term societal consequences of AI systems, especially in early- and mid-career stages where hiring algorithms may shape access to employment and mobility opportunities. Rather than defining youth narrowly by age, eligibility criteria were kept intentionally broad: participants needed only to not be AI experts and to identify with a group potentially subject to discrimination in hiring. This flexible framing aimed to centre lived experience and inclusivity over disciplinary credentials.
Ten individuals registered in advance. The group represented a wide range of disciplinary and professional backgrounds—including Business, Law, Computer Science, Engineering, Political Science, and work in the third sector—and included participants from both the Global North and Global Majority. They self-identified across a variety of gender identities and generational categories. This demographic diversity was essential to the project’s goal of surfacing intersectional perspectives on equity in algorithmic systems. Recruitment was conducted in collaboration with a youth-focused civic organisation, and was promoted via LinkedIn and Instagram by both the host and the NGO. Unlike the earlier workshop with AI practitioners, where all discussions were recorded, the youth workshop adopted a more cautious approach to documentation: only the final group presentations were recorded, creating a safer environment in case sensitive or personal experiences emerged.

\subsection{Pre-Workshop design}

A registration form  helped inform the facilitation plan and ensure demographic diversity within small groups. In addition, express consent was required on the registration form to collect the demographic data. Participants were asked to share:(i) basic demographics (e.g., gender identity, generation), (ii) self-assessed AI expertise (on a scale from 1 to 5), (iii) professional background, (iv) whether they belonged to a vulnerable group or marginalised group (e.g., based on migration status, race, disability, gender identity, socioeconomic background). These categories were self-defined by participants and helped situate how algorithmic systems may differently impact individuals depending on intersecting structural conditions, (v) their definition of equity. The registration form asked for an identifier that could be anything from initials, to colours, to nicknames that the participants needed to remember as they would be divided by groups according to the identifier provided. In addition, no email or contact information requested.

This data helped anticipate group dynamics and guided the introductory framing. Inspired by critical pedagogy \cite{freire2020pedagogy}, the workshop functions as both a site of collective inquiry and an educational space, aimed at building AI literacy as a form of civic empowerment. The facilitator developed a 10-minute presentation covering foundational concepts targeting a basic AI literacy: the hype around AI, technosolutionism, ethics-washing \cite{wagner2018ethics}, and algorithmic bias \cite{UNESCOBlush}. These ideas were illustrated with real-world case studies, such as AI tools that delivered discriminatory job ads based on gendered assumptions \cite{AlgorithmWatch}. To balance critique with possibility, the presentation also introduced a chatbot designed by a South African civil society organisation using participatory methods \cite{GRIT}. To avoid framing AI systems solely in negative or deterministic terms, the workshop introduced “Question Zero” \cite{DignumQuestionzero}—an ethical prompt that asks: Is it necessary to deploy an AI system in this context at all? This framing encouraged participants to step back from assumptions about innovation and consider whether algorithmic intervention was justified in the first place. As a conclusion in the presentation, the explanation about how initiatives such as the one attended by the participants are essential to turn community engagement into empowerment and how this can be beneficial to society. Drawing on the principles of design justice \cite{costanza2020design}, the workshops seek to redistribute not only access to technology but also the power to shape its development—foregrounding lived experience from insiders and outsiders of the AI lifecycle as central to ethical and equitable design processes. This framed participatory research not merely as data collection, but as a pathway to community empowerment, in other words, how groups that do not usually participate in the design, deployment and evaluation of AI systems can share their concerns and demands.

The workshop was developed in collaboration with a youth-led association, an NGO known for its sustained engagement with young people on issues of civic participation, human rights, and education. The organisation welcomed the opportunity to incorporate the workshop within its broader dissemination and outreach activities. They actively promoted the event via its social media channels and website, and played a crucial role in securing a venue. The workshop was hosted at a partner research center affiliated with the NGO, which provided the space free of charge. It is also essential to note that none of the participants, nor the association itself, received financial compensation for their involvement. 

While no identifying contact information was collected, participants were invited to express interest in future collaboration during the workshop itself. One participant subsequently volunteered to co-facilitate the next session. This informal pathway allowed for ongoing engagement without compromising privacy or requiring the collection of sensitive data. Also, it was agreed that the youth-based association would share on their social media the future works produced with the workshop.

\subsection{Workshop Agenda}

The workshop was structured as follows:

\begin{table}[h]
\centering
\begin{tabular}{|p{2.5cm}|p{3.5cm}|p{7.5cm}|}
\hline
\textbf{Time} & \textbf{Session} & \textbf{Objective} \\
\hline
10 min & Intro & Introduce AI literacy concepts \\
\hline
15 min & AI Practice & Exchange and consolidate ideas about AI \\
\hline
5 min & Debriefing & Discuss the results of the AI practice \\
\hline
15 min & Equity & Develop understanding of equity \\
\hline
5 min & Break & Coffee and pastry break \\
\hline
50 min & Indicators & Co-create potential indicators covering who, what, how, and why \\
\hline
15 min & Presentations & Groups present their work and facilitate a group-wide discussion \\
\hline
\end{tabular}
\caption{Workshop Agenda: “How to Think About Equity in the AI Ecosystem?”}
\label{tab:workshop-agenda}
\end{table}

The AI practice activity was conceived as an icebreaker using ideation cards \cite{ideationcards} to develop a deeper understanding of key concepts and ethical questions around AI systems.

\subsection{Activity 1: Understanding Equity}

The first major activity deepened participants’ understanding of equity by testing the working definition proposed by \cite{munarini2024equitable}:
'Providing meaningful access to the necessary resources for individuals who need it to belong to a community' in recruitment contexts.  Before the start of the activity,participants would be divided into small groups (3–7 people). The task would unfold in two stages:
1) Individual brainstorming (5 min): Participants wrote examples from recruitment processes on post-its, noting whether each was equitable, the identity categories involved (e.g. homeless, international student), and whether the example applied to them personally.
2) Group sharing (10 min): Teams placed their post-its along the equity spectrum and discussed how these examples illustrated or challenged the definition. 
Using large A1 sheets representing a spectrum from “no equity” to “equity,” participants began with individual brainstorming: they wrote examples from hiring processes on post-its—such as unpaid internships, algorithmic filtering of foreign names, or accommodations for disability disclosures. These examples included the identity groups impacted (e.g., international students, people with care responsibilities) and whether they personally related to them. In group discussions, post-its were mapped onto the equity spectrum to spark dialogue about what the definition captures—and where it falls short. This activity served as a warm-up for the more complex co-design phase that followed.
This activity was designed not only to surface concrete insights into equity in recruitment processes but also to prime participants for the more complex design thinking of the second half where AI systems would be included in the process.

\subsection{Activity 2: Co-designing equity indicators}

Building on the working definition of equity, the second activity focused on co-designing indicators—defined as “points of action” that can be used to evaluate whether an AI system supports equitable outcomes. Using a fictional AI hiring system called ARIA deployed by the recruitment company La Dolce Vita, developed specifically for this workshop as a composite example inspired by the Amazon case \cite{dastin2022amazon}. Working with the fictional AI hiring tool ARIA, participants used the prompt format “A [person/role] should [action]” and linked each indicator to a phase in the hiring pipeline, including a “Question Zero” stage that asks: Why use AI in this context at all? (e.g., “HR manager should review rejected applications flagged by the system”). Each indicator was attached to one or more phases of the AI hiring pipeline \cite{fabris2025hiringpipeline}, see Figure 1. Examples of indicators included:

“The HR could know more about the logical chain of the background of the position and then use this to screen candidates with better keywords”

“Hiring manager should prepare a set of questions that show applicants that the company really cares about diversity”

“Candidates should be given the possibility to do the same interview in different formats”

“Hiring platforms should clearly explain  how gamification assesses learning ability (after acquiring the job)”

Each group also discussed who the indicators targeted (e.g., recruiter, developer, platform owner), what they revealed (e.g., transparency, bias, oversight), and where they fit in the recruitment process (e.g., sourcing, shortlisting, final selection).

This activity bridged abstract ethical concerns and practical accountability tools, empowering participants to contribute to real-world design logics—even without technical backgrounds.

\section{Day of the workshop}

The workshop was scheduled to begin with ten registered participants. However, at the designated start time, only one person was present. This moment marked a critical inflection point in the facilitation process: whether to proceed, postpone, or adapt. Opting for the latter, the organiser extended the welcome period. Within the next forty minutes, three additional individuals joined—two of whom had not formally registered.
The final group consisted of four participants, each with distinct academic and professional trajectories. Though smaller than anticipated, the group composition offered an opportunity for a more intimate and dialogic mode of engagement. Accordingly, the original facilitation plan—structured around small-group breakout activities and collective synthesis—was adapted in real time to accommodate a single, continuous group dialogue. Their differing life experiences and positionalities shaped the discussion of equity and AI in rich and grounded ways—illustrating that epistemic value does not scale linearly with participant numbers.

There were four main positive outcomes from this new scenario. First, participants were able to speak at length about their personal experiences, professional aspirations, and perceptions of AI-driven systems. Second, the icebreaker and two activities were repurposed as group conversations rather than segmented exercises, which increased engagement. Third, the facilitator was able to provide tailored clarification of concepts and create space for affective responses around AI systems. Fourth, in comparison to the other workshops that hold different groups, it was easier to take observational notes.

Importantly, the absence of audiovisual recording—save for final presentations—enhanced the perceived safety of the space, encouraging openness and vulnerability. While this session did not meet expectations in terms of participant numbers, it arguably exceeded them in terms of epistemic richness and relational depth. The experience reinforced the importance of flexibility in participatory research design and highlighted the value of micro-scale engagements in surfacing contextually grounded insights.

\begin{figure}
    \centering
    \includegraphics[width=0.7\textwidth]{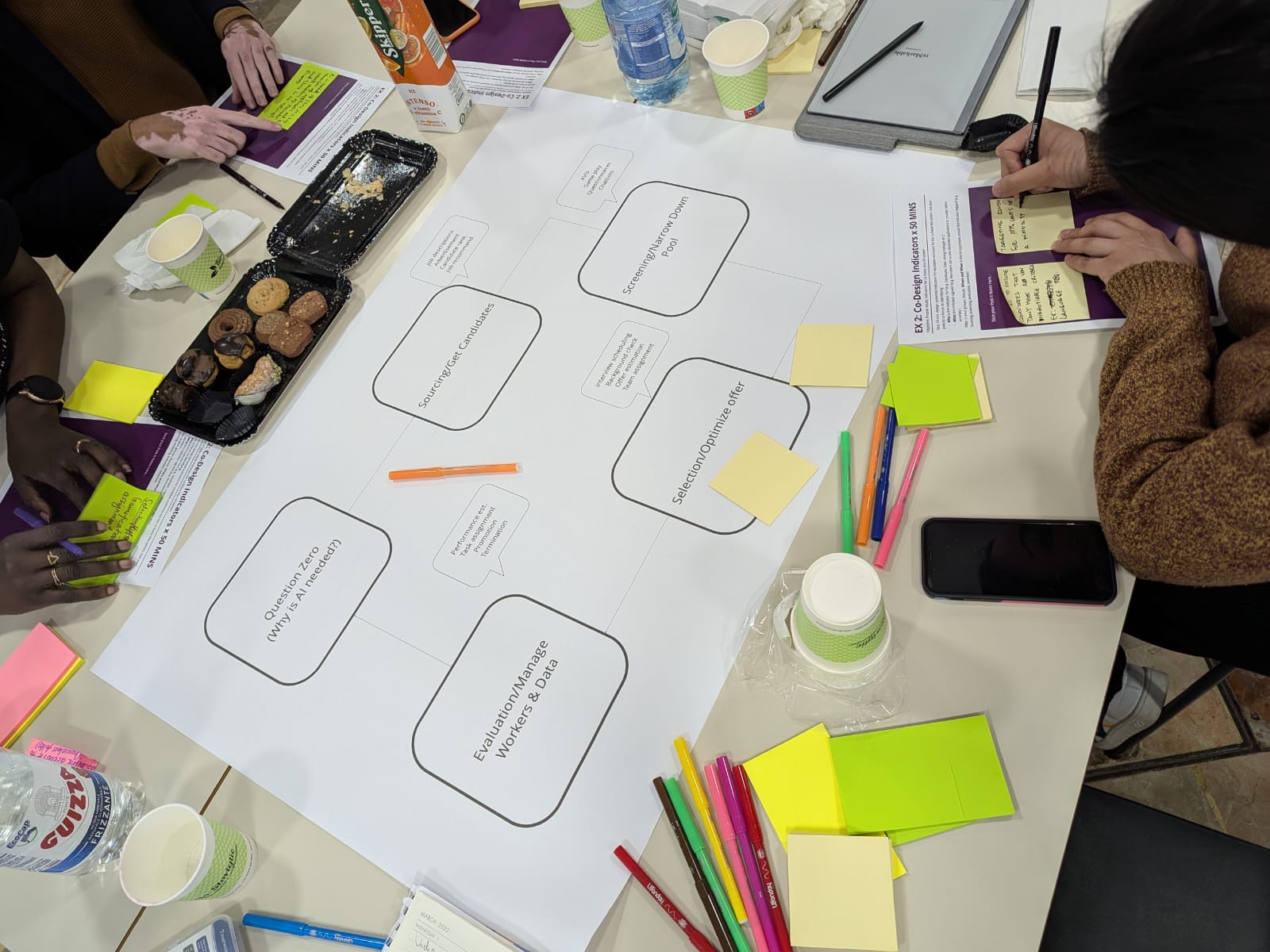}
    \caption{Participants working on equity indicators.}
    \label{act2}
\end{figure}

\section{Strategies for future workshops}

While the second workshop experienced significant initial challenges in participant turnout, it ultimately fulfilled its core aim: fostering a co-design space where diverse, non-expert voices could engage critically with AI systems used in recruitment. Despite beginning with only one attendee and facing the possibility of cancellation, the group eventually grew to four participants—two of whom were not previously registered. Importantly, this small group still embodied a range of gender identities, professional backgrounds, and geographies (Global North and Global Majority), aligning with the diversity goals outlined in the registration process. This diversity enriched the discussions and demonstrated that even with modest numbers, meaningful deliberation and learning can emerge when intersectional perspectives are present.
One key factor that contributed to the workshop’s success was the collaborative spirit of the participants and their interest in the topic of AI and equity. The active engagement of these individuals highlighted the value of designing workshops that prioritise inclusive participation and accessible language, especially when working with non-expert audiences. 

Reflecting on the logistical and organisational dimensions, several strategic insights emerged for ECRs aiming to replicate similar initiatives under resource constraints:

\begin{itemize}
    \item{Embed the workshop within a larger event: Hosting a workshop as part of a broader conference, symposium, or institutional programme significantly improves visibility, accessibility, and logistical ease. It also typically ensures access to basic infrastructure—such as a venue and refreshments—which can reduce both cost and coordination burden. More importantly, such contexts often come with a pre-registered audience, which increases the likelihood of turnout and offers opportunities for serendipitous engagement. All the other workshops followed this strategy and had a significant higher amount of participants.}
    \item {Build on existing networks and collaborations: Establishing connections with practitioners in participatory research in the AI ecosystem early in the design process proved invaluable. These connections provided not only critical feedback on the workshop structure but also offered opportunities for mutual learning and idea exchange. Informal conversations with professionals and academics helped refine the framing of the activities and the case study, and created pathways for further collaboration.
    \item {Leverage prior experience to build facilitation capacity: The development of the workshop was also rooted in the facilitator’s prior involvement in other co-design initiatives. Experience gained through participating in workshops in different projects helped build the confidence, adaptability, and methodological toolkit needed to design and implement a workshop closely aligned with the aims of a doctoral research project. }
    \item{Design strategies to remind registered participants: As the registration form did not collected identifiable data such as email, participants could not be reminded about the event. Using tools such as iCalendar or anonymous RSVP systems may help reduce no-shows while maintaining participant privacy. In this project, the decision not to collect contact information—such as names or emails—was based on ethical and methodological concerns, particularly the aim of lowering the barrier for marginalised or privacy-conscious individuals to participate in critical conversations about technologies that may already surveil or profile them.}
}    
\end{itemize}

One particularly meaningful outcome emerged after the second workshop, when a participant expressed a strong interest in the topic and volunteered to co-facilitate the third workshop. This transition — from participant to co-facilitator — illustrates what Sandra Harding \cite{harding1991whoser} and Patricia Collins \cite{collins2000black} might describe as a redistribution of epistemic authority, where lived experience and situated knowledge are not only valued but become integral to the design and facilitation process itself. This shift challenges traditional researcher–participant hierarchies and aligns with the principles of feminist participatory research \cite{costanza2020design}, which emphasize reciprocity, mutual learning, and empowerment \cite{fine2004critical}. The participant’s previous engagement added depth to the facilitation process, and their familiarity with the activities helped foster a more inclusive and collaborative environment. This movement suggests the potential of participatory workshops as not just data collection tools, but also as spaces of capacity-building and knowledge co-production — where ownership of the process can extend beyond the researcher's initial vision.
Although small in scale, the workshop generated outcomes that extended beyond anecdote. Participants collectively surfaced indicators for equitable AI systems, critically examined real-world equity challenges in hiring platforms, and one participant transitioned into a co-facilitator role for the next session. These instances reflect measurable redistributions of epistemic authority and support the view that participatory research—when conducted inclusively and reflexively—can yield both conceptual insight and practical transformation, even under resource constraints.

\section{Conclusion}

This paper has presented a reflective account of a participatory workshop series designed to explore equity in the AI ecosystem, particularly in the context of recruitment systems. Framed as a methodological reflection, the paper shares the challenges, unexpected moments, and practical strategies involved in designing and facilitating co-design workshops with affected groups—specifically identified as likely to face long-term impacts of algorithmic decision-making.

A central lesson emerging from this experience is that diversity in participation matters more than the number of attendees. Even when turnout was lower than expected, the presence of participants from varied gender identities, geographies, and professional backgrounds created meaningful dialogue and contributed to a deeper exploration of equity. 

The experience also revealed how practical constraints—such as the absence of a larger hosting event or funding for participant recruitment—can significantly shape outcomes. Embedding participatory activities within broader events or institutional structures can ease logistical pressures, enhance visibility, and increase turnout. At the same time, building relationships with practitioners and drawing on previous facilitation experience emerged as essential strategies to strengthen both the workshop design and delivery.
This paper does not claim to offer a universal model for ECRs working with participatory design in AI governance. Rather, it seeks to contribute to the growing body of work that takes participatory research seriously as both method and ethos. By sharing a grounded account of what worked, what did not, and how small-scale experiences can still generate valuable insights, this reflection aims to support other early-career researchers navigating similar paths—especially those working across disciplinary boundaries and without guaranteed institutional support.
In doing so, it reframes workshop ‘failures’ not as dead-ends, but as part of a productive learning process, where reflective practice itself becomes a mode of knowledge production. Co-designing equitable AI systems is not only a socio-technical challenge, but a relational and situated one—and participatory workshops, even with their uncertainties, remain a vital space for this work to unfold.

\bibliography{reflectivepaperworkshop}




\end{document}